\documentclass[sigconf, nonacm]{acmart}

\newcommand\vldbyear{2026}
\newcommand\vldbworkshop{Applied AI for Database Systems and Applications (AIDB 2026)}
\newcommand\vldbauthors{\authors}
\newcommand\vldbtitle{\shorttitle} 
\newcommand\vldbavailabilityurl{https://github.com/DataManagementLab/BespokeCard}
\newcommand\vldbpagestyle{plain}

\usepackage{hyperref}
\usepackage{cleveref}
\usepackage{caption}
\usepackage{graphicx}
\usepackage{multirow}
\usepackage{booktabs}
\usepackage[table]{xcolor}

\newcommand{\system}{Bespoke-Card}

\begin{document}
\title{Bespoke-Card: Why Tune When You Can Generate? Synthesizing Workload-Specific Cardinality Estimators}

\author{Johannes Wehrstein}
\orcid{0000-0002-7152-8959}
\affiliation{%
  \institution{
    Technical University of Darmstadt
  }
}
\author{Anton Winter}
\affiliation{%
  \institution{
    Technical University of Darmstadt
  }
}
\author{Timo Eckmann}
\orcid{0009-0007-7497-2389}
\affiliation{
  \institution{Technical University of Darmstadt}
}
\author{Carsten Binnig}
\orcid{0000-0002-2744-7836}
\affiliation{%
  \institution{
    Technical University of Darmstadt \& DFKI \& hessen.AI
  }
}

\begin{abstract}
Cardinality estimators are built to support arbitrary schemas and workloads, forcing them to rely on generic statistics even when the schema and workload is known in advance, leaving optimizers prone to large errors and poor plans. 
We present \system{}, an agent-driven system that synthesizes workload-specific cardinality estimators as executable code: a planning agent designs the estimators' strategies, a coding agent implements them, and a validator scores the estimates against true cardinalities and PostgreSQL estimates, forming a robust and deterministic harness.
Going beyond naive prompting, \system{} uses structured q-error feedback, regression analysis, concrete outlier subplans, a curriculum isolating join-only, filter-only, and full-subplan errors, and archival selection of the best implementation. 
Injecting its estimates into the optimizer cuts total PostgreSQL runtime on JOB by 33\% and reduces median q-error over all JOB subplans from 190.5 to 11.5 (-94\%), while synthesizing a strong estimator in under one hour for less than \$10.
\system{} is opening a new avenue for cardinality estimation next to classical generic estimators and learned estimator architectures.

\end{abstract}

\maketitle

\pagestyle{\vldbpagestyle}
\begingroup\small\noindent\raggedright\textbf{VLDB Workshop Reference Format:}\\
\vldbauthors. \vldbtitle. VLDB \vldbyear\ Workshop: \vldbworkshop.\\ 
\endgroup
\begingroup
\renewcommand\thefootnote{}\footnote{\noindent
This work is licensed under the Creative Commons BY-NC-ND 4.0 International License. Visit \url{https://creativecommons.org/licenses/by-nc-nd/4.0/} to view a copy of this license. For any use beyond those covered by this license, obtain permission by emailing \href{mailto:info@vldb.org}{info@vldb.org}. Copyright is held by the owner/author(s). Publication rights licensed to the VLDB Endowment. \\
\raggedright Proceedings of the VLDB Endowment. 
ISSN 2150-8097. \\
}\addtocounter{footnote}{-1}\endgroup

\ifdefempty{\vldbavailabilityurl}{}{
\vspace{.3cm}
\begingroup\small\noindent\raggedright\textbf{VLDB Workshop Artifact Availability:}\\
The source code, data, and/or other artifacts have been made available at \url{\vldbavailabilityurl}.
\endgroup
}

\vspace{-1ex}
\section{Introduction}
\vspace{-.5ex}

\begin{figure}
  \centering
  \includegraphics[width=\linewidth]{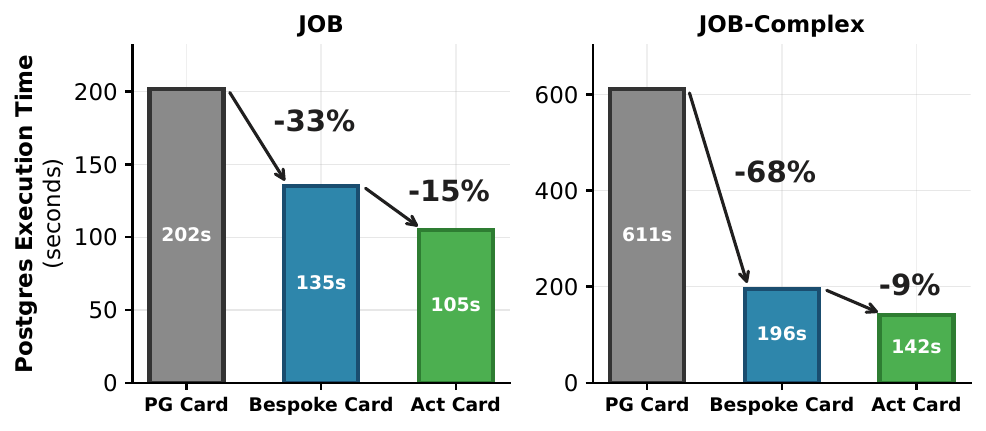}
  \vspace{-5ex}
  \caption{Bespoke-Card reduces total PostgreSQL runtime by 33\% on JOB and 68\% on JOB-Complex when its estimates are injected into the optimizer, compared to PostgreSQL using its own generic cardinality estimator.}
  \label{fig:total_e2e}
  \vspace{-4ex}
\end{figure}

\noindent\textbf{Cardinality estimation drives plan quality.}
Cardinality estimation is a central bottleneck in cost-based query optimization.
Query optimizers choose join orders, access paths, and physical operators based on estimates of intermediate result sizes. 
When these estimates are inaccurate, the optimizer may select plans that are orders of magnitude slower than alternatives. 
\citet{leis2015good} showed that such errors are not rare corner cases: on realistic join workloads, cardinality errors are often the dominant cause of poor plans, while cost models and enumeration strategies matter comparatively less once accurate cardinalities are available~\cite{leis2015good,DBLP:conf/cidr/LeisRGK017}.

\noindent\textbf{Generic statistics trade accuracy for generality.}
The difficulty is that real data violates the assumptions that make generic cardinality estimation tractable.
Value distributions are skewed, predicates are correlated, join-factors depend on filter literals, and many intermediate results are sparse or empty. 
Classical DBMS estimators address this problem with compact statistics such as histograms, most-common-value lists, samples, distinct-value counts, and coarse correlation statistics\cite{garcia2008database}. 
These statistics are cheap, robust, and easy to maintain, which explains their continued use in production. 
However, they are designed to work across arbitrary schemas, workloads, and data distributions. 
This generality comes at a cost: the estimator cannot fully exploit the concrete predicates, join paths, value domains, and correlations that characterize a specific workload.

\noindent\textbf{Learned estimators are limited.}
Learned cardinality estimators reduce some of this burden, but introduce different tradeoffs (see \Cref{sec:background} for a detailed discussion).
Query-driven approaches learn from query/cardinality pairs and can adapt to a workload, but require expensive label collection and retraining when workloads or data change. 
Data-driven approaches model the database distribution more directly, but still rely on a fixed model family and inference procedure chosen by system designers in advance. 
In both cases, the central design decision remains fixed: the estimator architecture is manually chosen, and only its parameters are adapted to the underlying database and workload.
Many learned estimators also support only simple queries, and commonly cannot even support queries of the complexity of the Join-Order-Benchmark (JOB)\cite{leis2015good}.
This limits their applicability in practice.

\noindent\textbf{Synthesizing the estimator itself.}
This paper explores a different point in the design space: can we synthesize the cardinality estimator itself for a fixed database and tune it for a workload?
We present \emph{\system{}}, a system that treats cardinality estimation as program code synthesis.
Given a database and workload, \system{} generates a cardinality estimator as code. 
The generated artifact first collects statistics on the database and leverages them to estimate the cardinality of an incoming query.
Thus, \system{} does not tune existing systems, produce optimizer hints, natural-language recommendations, or a trained model checkpoint. 
In contrast, it produces an executable estimator whose statistics and estimation logic are fully tailored to the declared database and workload.

\noindent\textbf{A database-and-workload contract.}
The database-specific scope of the estimator is intentional.
In practice the dataset is typically known, and the workload is often known, repeated, or explicitly declared\cite{van2024tpc}.
Together they form a contract: it exposes the data distribution, predicates, and join structure that the estimator may exploit, while freeing it from generality that is unnecessary for the task at hand.
This does not assume a fully static database: even under a non-negligible fraction of updates\cite{van2024tpc}, the schema and dominant query templates typically remain stable, and only the statistics depending on changed data need to be refreshed (\Cref{sec:overview},~\Cref{sec:conclusion}).
Crucially, specializing to this contract does not narrow the supported queries: even though an estimator is fixed to one dataset and tuned for one workload, it supports arbitrary SPAJ queries over that database (\Cref{sec:approach}).
For instance, in \Cref{sec:evaluation} we analyze estimators synthesized for the JOB \cite{leis2015good} and JOB-Complex\cite{wehrstein2025jobcomplex} benchmarks on the IMDb database.
Overall, \system{} does not attempt to replace a general-purpose estimator for arbitrary databases.
It exploits the dataset and workload knowledge that such estimators must ignore.

\noindent\textbf{Guiding synthesis with structured feedback.}
Synthesizing such an estimator, however, is not a matter of issuing a single prompt to a coding model.
Cardinality estimation combines statistical modeling with optimizer-facing constraints: filter and join errors interact multiplicatively, aggregate q-error may hide localized regressions, and improving difficult outliers can be less useful than fixing cases where simple additional statistics can help. 
\system{} therefore uses a structured synthesis loop. 
A planning agent proposes workload-specific statistics and estimation strategies, a coding agent implements them as executable code, and a deterministic evaluator measures each candidate against true cardinalities and a PostgreSQL baseline. 
The loop iteratively repairs the estimator using measured feedback rather than relying on the model's own assessment of quality.

\noindent\textbf{Reducing runtime.}
We evaluate \system{} on JOB\cite{leis2015good} and JOB-Complex\cite{wehrstein2025jobcomplex} by comparing generated estimates against PostgreSQL estimates over intermediate subplans and by measuring end-to-end runtime when the generated estimates are injected into PostgreSQL's optimizer. 
\system{} reduces total runtime by 33\% on JOB and 68\% on JOB-Complex as shown in \Cref{fig:total_e2e}. 
These results suggest that workload-specific executable synthesis can improve cardinality estimation accuracy, especially on workloads where generic assumptions fail.

\noindent This paper makes the following contributions:

\begin{enumerate}
  \vspace{-2ex}
  \setlength\itemindent{0pt}
  \setlength\leftskip{-15pt}
  \item \textbf{Executable cardinality-estimator synthesis.} 
We formulate cardinality estimation as the synthesis of a runnable estimator for a fixed database, tuned for a specific workload, producing executable statistics-generation and estimation code.

\item \textbf{A structured agent loop for estimator construction.} We introduce a planner/coder architecture in which statistic design and estimator implementation are separated and guided by deterministic evaluation feedback.

\item \textbf{Structured feedback for estimator improvement.} 
We design an evaluation harness that identifies pain-points and turns subplan-level estimation errors and regressions against PostgreSQL into structured repair signals for the synthesis loop.

\item \textbf{Curriculum optimization over subplans.} 
We show how join-only, filter-only, and full-subplan phases decompose the cardinality-estimation objective and help the coding agent localize estimation failures.

\item \textbf{Empirical study on JOB and JOB-Complex.} 
We show that synthesized estimators outperform PostgreSQL cardinality estimates and translate these improvements into lower end-to-end query runtimes on challenging join workloads.
\end{enumerate}
\vspace{-1ex}
\section{Background and Related Work}
\vspace{-.5ex}
\label{sec:background}

Cardinality estimation drives cost-based optimization, as the choice of join order, access paths, and physical operators depends on estimated intermediate-result sizes.
Cardinality estimation is hence a field of long-standing interest, with a large body of work on classical and learned approaches.

\noindent\textbf{Classical cardinality estimation.}
Classical cardinality estimators maintain compact base-data statistics such as histograms, most-common-value lists, distinct-count summaries, sketches, and samples \cite{selinger1979access,DBLP:conf/vldb/PoosalaI97,poosala1996improved,cormode2011synopses}, with sampling and join synopses adding robustness for some predicates and joins \cite{acharya1999join}.
Such summaries are cheap, predictable, and deeply integrated into production DBMSs, but they compress high-dimensional data and rely on assumptions like uniformity, independence, and containment.
They thus struggle with correlations, skew, complex predicates, and join-crossing dependencies, the failure modes exposed by realistic optimization benchmarks \cite{leis2015good,DBLP:conf/cidr/LeisRGK017}.

\noindent\textbf{Learned cardinality estimation.}
Learned estimators replace or augment hand-designed statistics with learned models.
Query-driven approaches map query features to cardinalities from labeled workloads, with MSCN modeling tables, joins, predicates, and samples via a set-convolution architecture \cite{kipf2019learned}.
Data-driven approaches instead model the data distribution: DeepDB learns relational sum-product networks without query labels \cite{hilprecht2019deepdb}, and NeuroCard uses autoregressive density models to capture cross-attribute and cross-table correlations \cite{yang2020neurocard}.
Hybrid and optimizer-aware variants such as UAE combine data- and query-driven signals \cite{wu2021unified}, while zero-shot and pretrained estimators like Iris and PRICE amortize learning across databases to cut per-database training cost \cite{lu2021iris,zeng2024price}, a direction pushed further by foundation database models \cite{wehrstein2025foundation}.
Although learned estimators can be more accurate in some cases, two obstacles have limited their adoption.
First, they are costly at both training and inference time: they require collecting labeled workloads and retraining whenever the data or workload changes, and their neural forward pass can be more expensive than the lightweight statistic lookups of classical or synthesized estimators.
Second, and more decisive in practice, available learned estimators support only numeric equality and range predicates, not the string/\texttt{LIKE}, null-sensitive, and correlated predicates that dominate realistic workloads such as JOB and JOB-Complex, rendering them inapplicable to the workloads we target and precluding a direct comparison (\Cref{sec:eval-setup}).
Hence, classical cardinality estimators are still the predominant choice in production DBMSs, and learned estimators have yet to see widespread adoption.

\noindent\textbf{Cardinality estimation as synthesis.}
\system{} occupies a different point in this space: rather than another fixed estimator architecture or a neural model trained over labels or samples, it synthesizes a workload-specific estimator as executable code.
This follows the broader Bespoke DBMS vision, in which workload-specific systems shed general-purpose overhead but require structured synthesis with validation and incremental refinement \cite{stonebraker2005onesize,wehrstein2026bespoke,eckmannfuture}.
Where Bespoke-OLAP \cite{wehrstein2026bespoke} targets full analytical engines, \system{} targets a single optimizer component: a planner designs statistics, a coder implements the estimator, and a deterministic evaluator gives feedback over q-error, regressions, and outlier subplans.
The contribution is thus not a model family, but an executable synthesis loop for cardinality estimation.
\vspace{-1ex}
\section{\system{} Overview}
\vspace{-.5ex}
\label{sec:overview}

\begin{figure*}
  \centering
  \includegraphics[width=.8\linewidth]{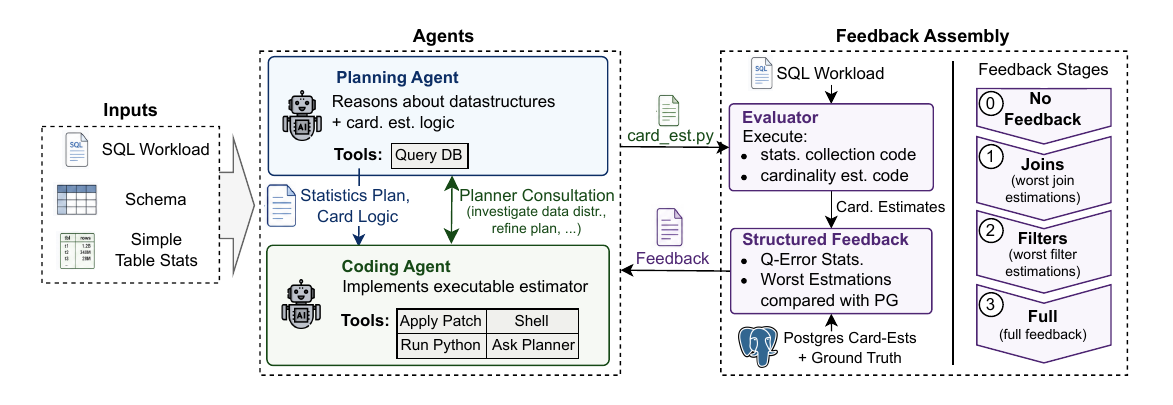}
  \vspace{-4ex}
  \caption{\system{} synthesizes a cardinality estimator given a dataset and tunes it for a given workload.
  Two agents, a planner and a coder, collaborate to generate the estimator using a closed-loop synthesis process.
  Different feedback stages (\textcircled{1}-\textcircled{3}) steer the synthesis loop to focus on specific estimation errors. The resulting estimator is an executable artifact producing cardinality estimates for SPAJ requests over the target database.}
  \label{fig:approach}
  \vspace{-1ex}
\end{figure*}

In this section, we first describe an overview of \system{} and then discuss the scope of this work.

\vspace{-.5ex}
\subsection{Synthesizing Bespoke Cardinality Estimators}
\label{sec:overview-synthesis}
The central idea is that \system{} synthesizes a cardinality estimator for a target database, using a given workload to guide specialization.
\Cref{fig:approach} provides an overview of \system{}.
The input to \system{} is a cardinality-estimation contract.
This contract consists of two components: a target database and a workload.
The target database fixes the schema, data distribution, and domain over which cardinalities are estimated.
The workload specifies the query patterns for which the generated estimator should be optimized.
Given this contract, \system{} synthesizes an executable cardinality estimator: program code that constructs workload-relevant statistics and implements estimation logic over the target database.

The output of \system{} is an executable estimator artifact.
For example, it may include classical summaries, samples, sketches, join statistics, correction rules, or hand-written estimation logic produced by the synthesis process.
It can be invoked through a cardinality-estimation interface and can therefore serve as a replaceable component inside a query optimizer.

\noindent\textbf{The Workload Guides Optimization.}
The specified workload acts as an optimization target.
It reveals which tables, joins, predicates, literals, correlations, and query shapes are important enough to specialize for.
Measured runtime performance and q-error on this workload then serve as the optimization signal, guiding the LLM agent to iteratively refine the engine toward this target.
The generated engine can then materialize statistics and estimation logic tailored to the workload's recurring structures.

The generated estimator is not restricted to the workload queries: it accepts any request in the supported query class over the target database.
In our research prototype, this class consists of select-project-aggregate-join (SPAJ) queries with complex predicates, similar to JOB.
Queries close to the targeted workload are expected to benefit most from the synthesized statistics and estimation logic, whereas queries far from it are still estimated but may fall back to more generic, less accurate behavior.
However, our approach in general is not limited to a specific query class, and the supported class can be expanded as needed.

\noindent\textbf{Bounded Scope is the Source of Specialization.}
General-purpose estimators pay a generality tax: because their logic and statistics must remain broadly applicable across many schemas, workloads, and data distributions, they cannot exploit regularities that are highly predictive in one database and workload, such as recurring join paths, stable predicate columns, skewed literals, or workload-specific correlations.
\system{} explores a different point in the design space: once the target database is known and a representative workload is available, the estimator can specialize its statistics and estimation logic to the parts of the database and query space that matter most, for example summaries for frequently joined table groups, predicate-specific statistics for recurring filter columns, or correlations that would be too narrow for a general DBMS statistics catalog.
This enables \system{} to outperform general-purpose estimators (\Cref{fig:total_e2e}).
Thus, the bounded scope of \system{} is not a limitation but the mechanism that enables specialization: the database defines the estimator's domain and the workload defines where accuracy matters most.
Overall, the goal is not to replace a general-purpose DBMS estimator for all possible workloads, but to synthesize a substantially better estimator for a known database and an important workload region.

\noindent\textbf{Synthesis Under Empirical Feedback.}
The challenge of this approach is how to produce such an estimator.
\system{} addresses this through an agentic synthesis process combined with deterministic measurement infrastructure.
At a high level, the synthesis process proposes statistics and estimation logic, materializes them as executable code, evaluates the resulting estimator, and uses the measured errors to guide the next iteration.

The important design principle is that empirical measurement remains outside the agent.
The agents generate and revise estimator code, but they do not decide by themselves whether the estimator is good.
Instead, synthesized estimators are evaluated by a deterministic harness.
This turns cardinality estimation into an executable synthesis problem under structured feedback, rather than a free-form prompting task.

\noindent\textbf{Supporting Changes in Data, Schema and Workload.}
Because \system{} is specialized to a target database and optimized for a representative workload, changes to either input may require adaptation.
If the data changes while schema and workload remain stable, the estimator's code stays valid and only its statistics need to be rebuilt by re-running the offline collection step, which is far cheaper than full resynthesis; between such refreshes, stale statistics degrade estimates gradually rather than breaking correctness, since the synthesized fallbacks still produce valid estimates.
If the schema changes, the synthesis process must extend the estimator to cover the new schema elements.
If the workload changes substantially, the estimator still supports queries within the supported class but may no longer be optimized for them, in which case \system{} can be rerun.
A generated estimator is thus not a permanent universal component, but a workload-shaped artifact that can be regenerated or extended as the database and workload evolve.

\vspace{-.5ex}
\subsection{Scope of this Work}
\label{sec:overview-scope}

Our research prototype targets SPAJ queries.
The evaluated workloads include queries with SQL complexity at the level of JOB, including multi-way joins and complex predicates.
The current prototype emits Python code.
We choose Python because it makes generated estimators easy to inspect, modify, and evaluate during synthesis.
The focus of this paper is estimation accuracy and query-performance impact, not low-level inference efficiency.
A production implementation could target programming language such as C++ or Rust, or compile the synthesized estimation logic into the optimizer directly.
\vspace{-1ex}
\section{Approach}
\vspace{-.5ex}
\label{sec:approach}

This section describes how \system{} synthesizes an executable cardinality estimator for a declared database/workload contract.

\vspace{-.5ex}
\subsection{Planning Workload-Specific Statistics}
\label{sec:approach-planner}

The first stage of \system{} is statistic planning.
It receives the database schema, the workload SQL, simple statistics about the tables (table cardinalities and distinct-value counts), and controlled read-only access to the underlying database.
The planner analyzes the workload from the perspective of cardinality estimation.
It identifies columns that appear in predicates, columns that participate in joins, recurring join paths, key and foreign-key relationships, string predicates, null-sensitive predicates, and table-specific sources of skew.
It further accesses the underlying database to identify data distributions or complicated correlations.
Based on this analysis, it proposes which statistics should be collected during estimator setup.
Typical statistics include per-column histograms, top-$k$ value summaries, distinct-value summaries, samples, conditional statistics, join-key summaries, and fallbacks for unsupported or less common cases.
Further, it decides how these statistics should be combined to estimate the cardinality for a query.

The planner is responsible for deciding on the estimator's layout and curating all necessary implementation information for the coder, but it does not write code.
This separation is deliberate.
Statistics design and estimator implementation require different types of reasoning: the planner should reason globally about workload structure and data distributions, whereas the coder should reason locally about executable code and measured failures.
The planner's output is therefore best understood as a design hypothesis, not as an optimal design.
The feedback loop described below may later reveal that some proposed statistics are insufficient, unnecessary, or need to be combined differently.

\vspace{-.5ex}
\subsection{Coding the Estimator}
\label{sec:approach-coder}

The second stage is implementation.
The coder receives the planner's statistics plan and turns it into a concrete estimator implementation.
The generated artifact is a Python module containing a \texttt{card\_estimator} class that implements the cardinality-estimation interface.
The coder can patch this file, run lightweight shell commands and perform python syntax checks.
It can also ask the planner for clarification questions or more details.

\noindent\textbf{Removing SQL parsing from the synthesis problem.}
The estimator is not invoked with raw SQL, instead, SQL requests are translated by our framework into a structured estimation request containing: the list of tables and their aliases, the list of filters over these tables, and the list of joins between these tables.
This representation removes SQL parsing from the synthesis problem and focuses the generated code on cardinality estimation itself.
This interface supports arbitrary structured SPAJ-style estimation requests over the target database schema.

\noindent\textbf{Collecting Statistics according to the Planner's Design.}
Statistics collection is executed once before evaluation.
It is implemented by the coder according to the planner's design.
It scans the database (the csv files through the provided readers) and constructs the proposed statistics.
This might include, for example, histograms, value-frequency maps, samples, string summaries, join-key maps, correlation summaries, or workload-specific lookup structures.
This is in line with classical DBMS design, where statistics are built offline once and used online for estimation.

\noindent\textbf{Leveraging Statistics for Estimation.}
For the cardinality estimation, the implementation combines statistics according to the structure of the request.
For single-table requests, it might estimate predicate selectivities using the relevant base-table statistics.
For join requests, it might combine table cardinalities, join-key statistics, key/foreign-key relationships, and filter selectivities.
For more complex requests, it might apply fallback rules when the estimator lacks specialized statistics.
All this considerations are hardcoded in the estimator code.
Although the resulting code may contain workload-specific logic, nevertheless it is still an estimator over arbitrary SPAJ requests because of its structured input format and the implemented fallback strategies.

\noindent\textbf{Tool constraints separating agent roles.}
The coder's tool access is restricted to make this division operational.
It can edit the estimator code through a patch interface, inspect a limited set of files, run its generated python code to check for syntax errors, and ask the planner for clarification.
The planner can inspect data but cannot edit code.
The coder can edit code but cannot freely inspect data.
This role separation is enforced by tools rather than by prompt instructions alone.
As a result, \system{} turns the intended planner/coder decomposition into an actual constraint on the synthesis process.

\vspace{-.5ex}
\subsection{Feedback Generation}
\label{subsec:feedback}

The evaluator is the only empirical signal source in \system{}.
It runs the synthesized estimator on a set of subplans, compares each estimate against the true cardinality and the PostgreSQL estimate, and passes the resulting diagnostics back as structured feedback.
The true cardinalities and PostgreSQL estimates are computed once up front and are reused throughout synthesis.

\noindent\textbf{Comparing against PostgreSQL to keep the feedback actionable.}
True cardinalities alone tell the coder \emph{that} an estimate is wrong, but not whether the error is fixable: a large q-error on an inherently hard subplan looks the same as one the estimator should easily get right.
Contrasting every estimate with a mature optimizer separates these cases.
Where \system{} is worse than PostgreSQL, the error is almost certainly addressable, because a generic optimizer with all its simplifying assumptions can answer accurately.
Where both our synthesized estimator and PostgreSQL fail, the case is hard for any estimator and harder to chase.
This comparison is essential: it steers the synthesis loop toward errors it can repair more easily and prevents it from getting stuck early on intrinsically hard subplans.

\begin{table}[h]
\centering
\small
\begin{tabular}{@{}ll@{}}
\toprule
\textbf{Feedback Field} & \textbf{Contained Information} \\
\midrule
\texttt{q\_error\_percentiles} & \begin{tabular}[c]{@{}l@{}}summary accuracy for \system{}\\ and PostgreSQL\end{tabular} \\\hline
\texttt{total\_regression\_rate} & \begin{tabular}[c]{@{}l@{}}fraction of subplans where \system{}\\  is worse than PostgreSQL\end{tabular} \\\hline
\texttt{grouped\_q\_error} & \begin{tabular}[c]{@{}l@{}}accuracy grouped along a feedback-\\ stage-specific dimension\end{tabular} \\\hline
\texttt{grouped\_regression\_rate} & \begin{tabular}[c]{@{}l@{}}regression rate grouped along \\ feedback-stage-specific dimension\end{tabular} \\\hline
\texttt{outliers} & \begin{tabular}[c]{@{}l@{}}10 worst overestimates and under-\\ estimates with full subplan context\end{tabular} \\\hline
\texttt{estimator\_size} & memory footprint of statistics \\
\bottomrule
\end{tabular}
\caption{Schema of the structured feedback. Specific information (e.g. the aggregation dimension of q-error and regression rate) is set per feedback-stage (\Cref{tab:feedback-stages}).}
\label{tab:feedback-fields}
\vspace{-5ex}
\end{table}

\noindent\textbf{Structured feedback.}
The feedback is structured as shown in \Cref{tab:feedback-fields}.
It is designed for \emph{improvement}, not only reporting: percentiles summarize overall quality, grouped diagnostics localize systematic failure modes e.g. over num-tables or filter types, and outliers give concrete failing subplans.
The outlier segment, contains the ten worst overestimates and underestimates based on the comparison with PostgreSQL as discussed previously.
They are annotated with the true cardinality, PostgreSQL estimate, generated estimate, and error direction (over/underestimation).
This gives the coder concrete cases to chase and a sense of the error's magnitude and direction, rather than just an abstract q-error number.
Alongside these concrete cases, the feedback reports the summed q-error over all outliers together with the count of over- and under-estimates among them, giving both specific signals for improvement and an overall trend of the tail's direction and magnitude.
The \texttt{estimator\_size} field reports the in-memory footprint of the collected statistics: if accuracy is bought with statistics that grow too large, it must trade precision for compactness, for example by coarsening histograms or dropping rarely-used per-column structures.

\noindent\textbf{Feedback stages steer optimization.}
During synthesis, \system{} evaluates the estimator in three feedback stages (\Cref{fig:approach}\textcircled{1}-\textcircled{3} ).
Each stage focuses on a different set of subplans and groups the diagnostics along a different dimension (\Cref{tab:feedback-fields}), isolating one source of error at a time.
The overall loop is still free to explore any change, but the feedback structure steers it toward changes that fix specific errors in a specific order: first joins, then filters, then their interaction.
This structured approach is more effective than a single feedback stage that mixes all subplans together, because it helps the coder to identify and fix specific errors rather than chasing an undifferentiated overall signal.

\begin{table}[h]
\centering
\small
\begin{tabular}{@{}llll@{}}
\toprule
\textbf{Stage} & \textbf{Subplans} & \textbf{Grouping} & \textbf{\begin{tabular}[c]{@{}l@{}}Isolated \\ error source\end{tabular}} \\ 
\midrule
1. Join & \begin{tabular}[c]{@{}l@{}}no-filter\\ subplans\end{tabular} & \begin{tabular}[c]{@{}l@{}}\#joined tables, \\ join structure\end{tabular} & join card. \\\hline
2. Filter & \begin{tabular}[c]{@{}l@{}}no-join\\ subplans\end{tabular} & \begin{tabular}[c]{@{}l@{}}table, column, \\ \#filters, predicate type\end{tabular} & \begin{tabular}[c]{@{}l@{}}base-table\\ selectivity\end{tabular} \\\hline
3. Full & all & \begin{tabular}[c]{@{}l@{}}subplan size, \\ predicate type\end{tabular} & \begin{tabular}[c]{@{}l@{}}filter--join\\ interaction\end{tabular} \\
\bottomrule
\end{tabular}
\caption{The three feedback stages. Each uses the schema of \Cref{tab:feedback-fields} on a different subset of plans. The feedback first focuses on repairing join cardinalities, then filter selectivities, then their interaction.}
\label{tab:feedback-stages}
\vspace{-5ex}
\end{table}

\noindent\textbf{Stage \textcircled{1}: join feedback.}
The first feedback stage covers only subplans with no filters.
Here, predicate selectivity does not contribute to the error, so the full signal is about join cardinalities: key/foreign-key behavior, join-hit rates, many-to-many joins, table-pair skew, and multiway join composition.
Grouping by number of joined tables and join structure lets the coder tune join factors (how many matching partners a tuple has in the other table) and the correlation/independence assumptions between tables.
This stage runs first, since join cardinalities can easily explode, making them a dominant source of the overall estimation error.

\noindent\textbf{Stage \textcircled{2}: filter feedback.}
The second stages covers subplans without joins, so the estimate depends only on table cardinalities and predicate selectivities.
Errors here point to histograms, most-common-value statistics, range handling, string predicates, null handling, correlated predicates, or fallbacks for unsupported predicate forms.
Because standard statistics should already handle many single-table cases, a PostgreSQL regression in this stage is a concrete signal that the estimator's workload-specific logic has broken baseline behavior rather than merely failed on a hard join.

\noindent\textbf{Stage \textcircled{3}: full-subplan feedback.}
The third stage covers all subplans of the workload to assess the interaction between predicate and join selectivity.
This is often where estimators fail: a filter changes join-hit rates, correlations span tables, and independent errors multiply across a join tree.
With joins and filters already addressed in isolation, a regression that appears only here implicates the interaction rather than either component alone.

\vspace{-.5ex}
\subsection{Archival and Candidate Selection}
\label{sec:approach-archive}

In general, every optimization stage improves the results, however we do still have checkpointing in place to recover from non-monotonic behavior of the agents, which can be caused by e.g. hallucinated information, misinterpretation of feedback, or bad repair suggestions.
Hence, archival is important for robustness, however rarely noticed in our experiments.

\vspace{-.5ex}
\subsection{Operational Constraints and Reproducibility}
\label{sec:approach-reproducibility}

Several additional constraints make the loop controlled and reproducible.
The evaluator runs outside the agents as a deterministic subprocess and communicates with them only through the structured feedback messages.
The coder edits only the generated estimator file.
The planner's database access is read-only and bounded.
Planner re-entry from the coder is allowed, but bounded, so that the coder can request design clarification without acquiring unrestricted data access.

\system{} also records resource usage for the synthesis process.
Token counts, tool invocations, response times, evaluation time, and phase-level costs are logged separately for reporting and improving the harness.
Together with archived artifacts and deterministic evaluation, this makes the final synthesized cardinality estimator an auditable artifact of a structured synthesis loop rather than an opaque output of a single model call.
\vspace{-1ex}
\section{Evaluation}
\vspace{-.5ex}
\label{sec:evaluation}

We evaluate \system{} to answer five questions.
First, does injecting the synthesized estimates into the optimizer translate into faster query execution (\Cref{sec:eval-e2e})?
Second, are the synthesized estimates actually more accurate than PostgreSQL's, and where does the improvement come from (\Cref{sec:eval-joins,sec:eval-qerror})?
Third, what strategies are implemented in the synthesized estimators (\Cref{sec:eval-semantic})?
Fourth, how much does the staged feedback loop contribute over the initial, feedback-free design (\Cref{sec:eval-stages})?
Fifth, what does \system{} cost to run, and how large are the artifacts it produces (\Cref{sec:eval-memory,sec:eval-cost})?

\vspace{-.5ex}
\subsection{Experiment Setup}
\label{sec:eval-setup}

\noindent\textbf{Hardware and DBMS.}
All experiments run on a machine with two Intel Xeon Gold 5220 CPUs (2.20\,GHz, $2\times18$ physical cores) and 504\,GB of RAM.
We use PostgreSQL~18, tuned for analytical workloads with PGTune\cite{kliukin2014pgtune}, apply multithreading with 8 workers, and create indexes on all primary keys.

\noindent\textbf{Workloads.}
We evaluate \system{} separately on two workloads over the IMDb dataset: the Join Order Benchmark (JOB\cite{leis2015good}) with 113 queries, and JOB-Complex\cite{wehrstein2025jobcomplex} with 30 queries.
Both are commonly used benchmarks for cardinality estimation and QO research, and represent realistic analytical workloads with complex join patterns and predicates.
We synthesize a separate bespoke estimator for each workload.

\noindent\textbf{Baselines and cardinality injection.}
We compare \system{} against two baselines.
The first is PostgreSQL's built-in estimator (\emph{estimated cards}), representing a mature, general-purpose statistics catalog.
The second injects the \emph{true} cardinalities (\emph{actual cards}) into the optimizer, representing the performance ceiling achievable through perfect cardinality estimation alone.
True cardinalities are obtained by running \texttt{EXPLAIN ANALYZE} on every query and subplan.
We inject both the true and the \system{} estimates into PostgreSQL's optimizer using PG-Lab\cite{DBLP:journals/pacmmod/BergmannHHL25}, leaving join enumeration, cost model, and physical operator selection unchanged.
Learned cardinality estimators are not included as a baseline.
To the best of our knowledge, available query- and data-driven models do not support \texttt{LIKE}/string and other complex predicates, and thus cannot handle real-world workloads such as JOB and JOB-Complex, whose predicates are heavily string-based.
This makes them inapplicable here and precludes a direct comparison.

\noindent\textbf{LLM Model.}
GPT-5.4 is used as the underlying LLM for both the planning and the coding agent, with the same model applied across all synthesis stages.

\vspace{-.5ex}
\subsection{End-to-End Runtime Impact}
\label{sec:eval-e2e}

We first measure the practical payoff: the total runtime of all queries in the workload when the PostgreSQL optimizer uses \system{}'s estimates, PostgreSQL's own estimates, and the true cardinalities.
\Cref{fig:total_e2e} reports the results.

The synthesized \system{} estimators of both workloads substantially reduce total runtime over PostgreSQL's default estimates and closes most of the gap to the unattainable true-cardinality ceiling.
On JOB, \system{} cuts total runtime by 33\% relative to PostgreSQL's estimates (from 202\,s to 135\,s). 
The remaining gap to the true-cardinality ceiling is under 15\% of PostgreSQL's original runtime.
On JOB-Complex, where generic assumptions break down more severely, the effect is larger: \system{} reduces total runtime by 68\% (from 611\,s to 196\,s). 
Compared to PostgreSQL's original runtime, the remaining gap to the true-cardinality ceiling (142\,s) is under 9\%.
These results show that the estimates of \system{} translate directly into better plans and faster execution, with the largest benefits on the workload where PostgreSQL's generic statistics fail most.

\vspace{-.5ex}
\subsection{Estimation Quality on Joins}
\label{sec:eval-joins}

\begin{figure*}[t]
  \centering
  \includegraphics[width=\linewidth]{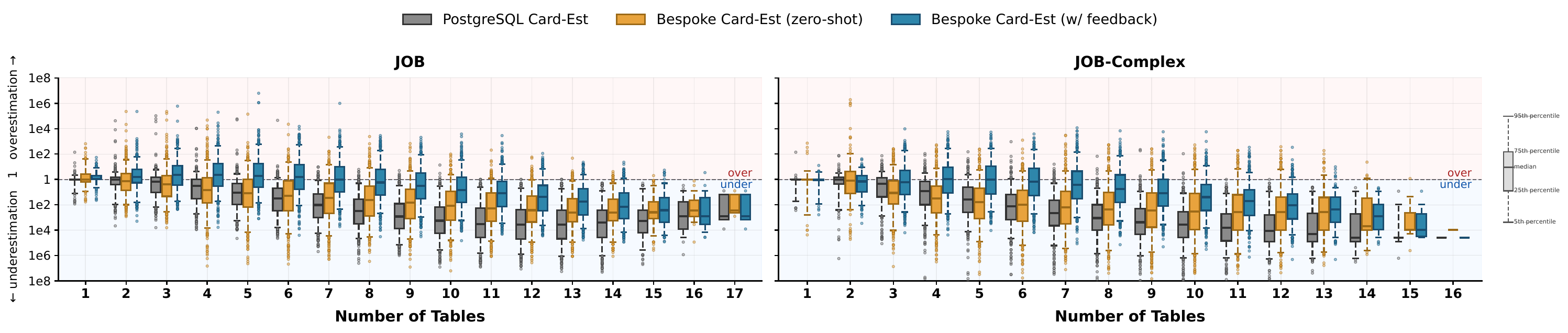}
  \vspace{-5ex}
  \caption{Quality of cardinality estimates for multi-join queries compared to the true cardinalities. 
  \system{} \textit{w/ feedback} and \textit{zero-shot} denote the performance with and without feedback.
  Each boxplot summarizes the error distribution over all subplans of a given size (across all queries in the workload). 
  \system{} provides more accurate estimates and exhibits a less pronounced underestimation trend compared to PostgreSQL.}
  \label{fig:error_over_tables}
  \vspace{-2ex}
\end{figure*}

To understand where the runtime improvements come from, we next examine estimation accuracy as a function of number of joined tables in \Cref{fig:error_over_tables}.
PostgreSQL exhibits the well-known underestimation trend on multi-join subplans\cite{leis2015good}: as more tables are joined, independence assumptions compound and estimates fall increasingly below the true cardinalities.
\system{} stays much closer to the true cardinalities across all subplan sizes.
A mild underestimation trend only becomes visible from roughly eight joined tables onward, and even there the errors remain considerably smaller than PostgreSQL's.
This behavior holds for both synthesized estimators (JOB and JOB-Complex), indicating that the synthesis approach produces robust join logic on both workloads rather than an effect specific to a single query set.

\vspace{-.5ex}
\subsection{Q-Error Distribution}
\label{sec:eval-qerror}

\begin{figure}
  \centering
  \includegraphics[width=\linewidth]{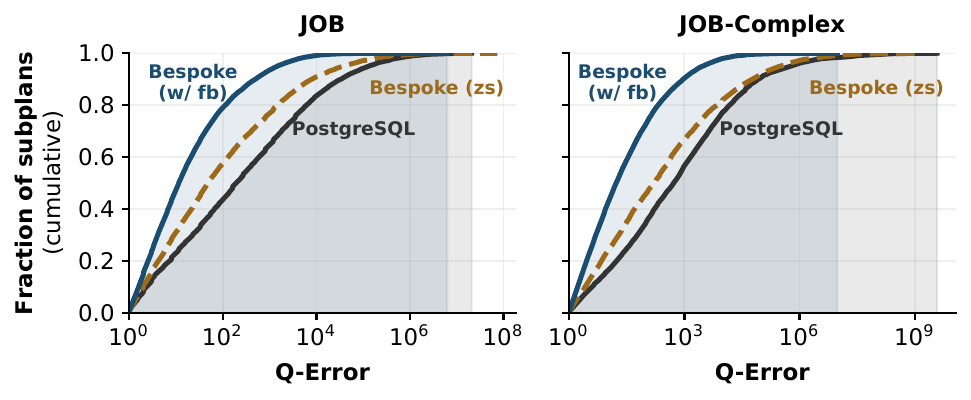}
  \vspace{-6ex}
  \caption{Q-error distribution of cardinality estimates compared to the true cardinalities, over all subplans of all queries in the workload. 
  \system{} \textit{w/ fb} and \textit{zs} describes the performance with and without feedback (zero-shot).
  \system{} achieves overall much better estimates than PostgreSQL.}
  \label{fig:q_error_distribution_all}
  \vspace{-1ex}
\end{figure}

\begin{table}[]
  \centering
  \scalebox{0.95}{
    \begin{tabular}{ccrrrrr}
      \toprule
      \textbf{Workload} & \textbf{System} & \textbf{q5} & \textbf{q25} & \textbf{q50} & \textbf{q90} & \textbf{q95} \\ \midrule
      \multirow{2}{*}{\textbf{JOB}} & \textit{PostgreSQL} & 1.50 & 12.14 & 190.5 & 33.7k & 132k \\
      & \textit{\system} & 1.24 & 3.21 & 11.5 & 0.5k & 1k \\
      \midrule
      \multirow{2}{*}{\textbf{\begin{tabular}[c]{@{}c@{}}JOB-\\ Complex\end{tabular}}} & \textit{PostgreSQL} & 1.87 & 32.97 & 530.4 & 83.0k & 590k \\
      & \textit{\system} & 1.29 & 3.82 & 16.9 & 0.9k & 3k \\
      \bottomrule
    \end{tabular}
  }
  \caption{Quantiles of the q-error distribution. \system{} achieves significantly more accurate estimates than PostgreSQL at every percentile.}
  \label{tab:qerrors}
  \vspace{-4ex}
\end{table}

We now consider the full q-error distribution over all subplans, summarized in \Cref{tab:qerrors} and visualized in \Cref{fig:q_error_distribution_all}.
\system{} improves estimation accuracy at every percentile and on both workloads.
On JOB, it reduces the median q-error from 190.5 to 11.5 (a 94\% reduction) and, more importantly, compresses the error tail: the 90th percentile drops from 33.7k to 0.5k and the 95th percentile from 132k to 1k, almost two orders of magnitude.
The improvement is even more pronounced on JOB-Complex, where PostgreSQL's median q-error of 530.4 already reflects the difficulty of the workload; \system{} brings it down to 16.9, with the 90th and 95th percentiles falling from 83.0k and 590k to 0.9k and 3k, respectively.
The full distribution in \Cref{fig:q_error_distribution_all} confirms this picture: \system{} shifts mass toward low q-errors and sharply thins the heavy tail that drives PostgreSQL's worst plan choices.

\vspace{-.5ex}
\subsection{Strategies in the Generated Estimators}
\label{sec:eval-semantic}

\begin{table}
  \centering
  \scalebox{0.88}{%
\definecolor{grp1560153e}{rgb}{0.941,0.976,1.000}
\definecolor{grpa5699a3f}{rgb}{0.941,0.992,0.957}
\definecolor{grpa0fcf967}{rgb}{1.000,0.969,0.929}
\begin{tabular}{l l c c c }
\hline
 & {\renewcommand{\arraystretch}{0.85}\begin{tabular}[c]{@{}c@{}}Bespoke and Non-\\[-0.4ex]Bespoke Card-Est. Strategies\end{tabular}} & {\renewcommand{\arraystretch}{0.85}\begin{tabular}[c]{@{}c@{}}Classical\\[-0.4ex]Exotic\end{tabular}} & {\renewcommand{\arraystretch}{0.85}\begin{tabular}[c]{@{}c@{}}JOB\end{tabular}} & {\renewcommand{\arraystretch}{0.85}\begin{tabular}[c]{@{}c@{}}JOB-\\[-0.4ex]Complex\end{tabular}} \\
\hline
\cellcolor{grp1560153e}  & \cellcolor{grp1560153e} {\renewcommand{\arraystretch}{0.85}\begin{tabular}[c]{@{}l@{}}Reservoir Sampling\end{tabular}} & \cellcolor{grp1560153e} {\renewcommand{\arraystretch}{0.85}\begin{tabular}[c]{@{}c@{}}C\end{tabular}} & \cellcolor{grp1560153e} {\renewcommand{\arraystretch}{0.85}\begin{tabular}[c]{@{}c@{}}100\%\end{tabular}} & \cellcolor{grp1560153e} {\renewcommand{\arraystretch}{0.85}\begin{tabular}[c]{@{}c@{}}100\%~cols\end{tabular}} \\
\cellcolor{grp1560153e}  & \cellcolor{grp1560153e} {\renewcommand{\arraystretch}{0.85}\begin{tabular}[c]{@{}l@{}}n-distinct Estimation\end{tabular}} & \cellcolor{grp1560153e} {\renewcommand{\arraystretch}{0.85}\begin{tabular}[c]{@{}c@{}}C\end{tabular}} & \cellcolor{grp1560153e} {\renewcommand{\arraystretch}{0.85}\begin{tabular}[c]{@{}c@{}}100\%\end{tabular}} & \cellcolor{grp1560153e} {\renewcommand{\arraystretch}{0.85}\begin{tabular}[c]{@{}c@{}}100\%~cols\end{tabular}} \\
\cellcolor{grp1560153e}  & \cellcolor{grp1560153e} {\renewcommand{\arraystretch}{0.85}\begin{tabular}[c]{@{}l@{}}Frequent Values plus\\[-0.4ex]Rare-Value Remainder\end{tabular}} & \cellcolor{grp1560153e} {\renewcommand{\arraystretch}{0.85}\begin{tabular}[c]{@{}c@{}}C\end{tabular}} & \cellcolor{grp1560153e} {\renewcommand{\arraystretch}{0.85}\begin{tabular}[c]{@{}c@{}}62\%\end{tabular}} & \cellcolor{grp1560153e} {\renewcommand{\arraystretch}{0.85}\begin{tabular}[c]{@{}c@{}}62\%~cols\end{tabular}} \\
\cellcolor{grp1560153e}  & \cellcolor{grp1560153e} {\renewcommand{\arraystretch}{0.85}\begin{tabular}[c]{@{}l@{}}Equi-Depth Histogram\end{tabular}} & \cellcolor{grp1560153e} {\renewcommand{\arraystretch}{0.85}\begin{tabular}[c]{@{}c@{}}C\end{tabular}} & \cellcolor{grp1560153e} {\renewcommand{\arraystretch}{0.85}\begin{tabular}[c]{@{}c@{}}4\%\end{tabular}} & \cellcolor{grp1560153e} {\renewcommand{\arraystretch}{0.85}\begin{tabular}[c]{@{}c@{}}55\%~cols\end{tabular}} \\
\cellcolor{grp1560153e}  & \cellcolor{grp1560153e} {\renewcommand{\arraystretch}{0.85}\begin{tabular}[c]{@{}l@{}}Frequent Value Pairs\\[-0.4ex]across Columns\end{tabular}} & \cellcolor{grp1560153e} {\renewcommand{\arraystretch}{0.85}\begin{tabular}[c]{@{}c@{}}E\end{tabular}} & \cellcolor{grp1560153e} {\renewcommand{\arraystretch}{0.85}\begin{tabular}[c]{@{}c@{}}--\end{tabular}} & \cellcolor{grp1560153e} {\renewcommand{\arraystretch}{0.85}\begin{tabular}[c]{@{}c@{}}34\%~cols\end{tabular}} \\
\cellcolor{grp1560153e}  & \cellcolor{grp1560153e} {\renewcommand{\arraystretch}{0.85}\begin{tabular}[c]{@{}l@{}}Join-Key Fan-out\\[-0.4ex]/ Coverage Stats\end{tabular}} & \cellcolor{grp1560153e} {\renewcommand{\arraystretch}{0.85}\begin{tabular}[c]{@{}c@{}}E\end{tabular}} & \cellcolor{grp1560153e} {\renewcommand{\arraystretch}{0.85}\begin{tabular}[c]{@{}c@{}}8\%\end{tabular}} & \cellcolor{grp1560153e} {\renewcommand{\arraystretch}{0.85}\begin{tabular}[c]{@{}c@{}}16\%~cols\end{tabular}} \\
\cellcolor{grp1560153e}  & \cellcolor{grp1560153e} {\renewcommand{\arraystretch}{0.85}\begin{tabular}[c]{@{}l@{}}Exact Stats for\\[-0.4ex]Tiny Tables\end{tabular}} & \cellcolor{grp1560153e} {\renewcommand{\arraystretch}{0.85}\begin{tabular}[c]{@{}c@{}}C\end{tabular}} & \cellcolor{grp1560153e} {\renewcommand{\arraystretch}{0.85}\begin{tabular}[c]{@{}c@{}}--\end{tabular}} & \cellcolor{grp1560153e} {\renewcommand{\arraystretch}{0.85}\begin{tabular}[c]{@{}c@{}}11\%~cols\end{tabular}} \\
\cellcolor{grp1560153e}  & \cellcolor{grp1560153e} {\renewcommand{\arraystretch}{0.85}\begin{tabular}[c]{@{}l@{}}Substring and Prefix\\[-0.4ex]Text Stats\end{tabular}} & \cellcolor{grp1560153e} {\renewcommand{\arraystretch}{0.85}\begin{tabular}[c]{@{}c@{}}C\end{tabular}} & \cellcolor{grp1560153e} {\renewcommand{\arraystretch}{0.85}\begin{tabular}[c]{@{}c@{}}6\%\end{tabular}} & \cellcolor{grp1560153e} {\renewcommand{\arraystretch}{0.85}\begin{tabular}[c]{@{}c@{}}9\%~cols\end{tabular}} \\
\cellcolor{grp1560153e}  & \cellcolor{grp1560153e} {\renewcommand{\arraystretch}{0.85}\begin{tabular}[c]{@{}l@{}}Precomputed Stats for\\[-0.4ex]Query Literals\end{tabular}} & \cellcolor{grp1560153e} {\renewcommand{\arraystretch}{0.85}\begin{tabular}[c]{@{}c@{}}E\end{tabular}} & \cellcolor{grp1560153e} {\renewcommand{\arraystretch}{0.85}\begin{tabular}[c]{@{}c@{}}3\%\end{tabular}} & \cellcolor{grp1560153e} {\renewcommand{\arraystretch}{0.85}\begin{tabular}[c]{@{}c@{}}9\%~cols\end{tabular}} \\
\cellcolor{grp1560153e}  & \cellcolor{grp1560153e} {\renewcommand{\arraystretch}{0.85}\begin{tabular}[c]{@{}l@{}}Info Values Conditioned\\[-0.4ex]on Info Type\end{tabular}} & \cellcolor{grp1560153e} {\renewcommand{\arraystretch}{0.85}\begin{tabular}[c]{@{}c@{}}E\end{tabular}} & \cellcolor{grp1560153e} {\renewcommand{\arraystretch}{0.85}\begin{tabular}[c]{@{}c@{}}3\%\end{tabular}} & \cellcolor{grp1560153e} {\renewcommand{\arraystretch}{0.85}\begin{tabular}[c]{@{}c@{}}5\%~cols\end{tabular}} \\
\cellcolor{grp1560153e}\multirow{-11}{*}{\rotatebox[origin=c]{90}{{\renewcommand{\arraystretch}{0.85}\begin{tabular}[c]{@{}c@{}}Statistics\\[-0.4ex]\& Synopses\end{tabular}}}} & \cellcolor{grp1560153e} {\renewcommand{\arraystretch}{0.85}\begin{tabular}[c]{@{}l@{}}Hand-Tuned Sample\\[-0.4ex]Sizes per Table\end{tabular}} & \cellcolor{grp1560153e} {\renewcommand{\arraystretch}{0.85}\begin{tabular}[c]{@{}c@{}}E\end{tabular}} & \cellcolor{grp1560153e} {\renewcommand{\arraystretch}{0.85}\begin{tabular}[c]{@{}c@{}}\ensuremath{\checkmark}\end{tabular}} & \cellcolor{grp1560153e} {\renewcommand{\arraystretch}{0.85}\begin{tabular}[c]{@{}c@{}}\ensuremath{\checkmark}\end{tabular}} \\
\hline
\cellcolor{grpa5699a3f}  & \cellcolor{grpa5699a3f} {\renewcommand{\arraystretch}{0.85}\begin{tabular}[c]{@{}l@{}}Floor Very Small\\[-0.4ex]AND-Filter Estimates\end{tabular}} & \cellcolor{grpa5699a3f} {\renewcommand{\arraystretch}{0.85}\begin{tabular}[c]{@{}c@{}}E\end{tabular}} & \cellcolor{grpa5699a3f} {\renewcommand{\arraystretch}{0.85}\begin{tabular}[c]{@{}c@{}}\ensuremath{\checkmark}\end{tabular}} & \cellcolor{grpa5699a3f} {\renewcommand{\arraystretch}{0.85}\begin{tabular}[c]{@{}c@{}}--\end{tabular}} \\
\cellcolor{grpa5699a3f}  & \cellcolor{grpa5699a3f} {\renewcommand{\arraystretch}{0.85}\begin{tabular}[c]{@{}l@{}}Run Predicates on\\[-0.4ex]Sampled or Exact Rows\end{tabular}} & \cellcolor{grpa5699a3f} {\renewcommand{\arraystretch}{0.85}\begin{tabular}[c]{@{}c@{}}E\end{tabular}} & \cellcolor{grpa5699a3f} {\renewcommand{\arraystretch}{0.85}\begin{tabular}[c]{@{}c@{}}--\end{tabular}} & \cellcolor{grpa5699a3f} {\renewcommand{\arraystretch}{0.85}\begin{tabular}[c]{@{}c@{}}\ensuremath{\checkmark}\end{tabular}} \\
\cellcolor{grpa5699a3f}  & \cellcolor{grpa5699a3f} {\renewcommand{\arraystretch}{0.85}\begin{tabular}[c]{@{}l@{}}Info Filters Conditioned\\[-0.4ex]on Info Type\end{tabular}} & \cellcolor{grpa5699a3f} {\renewcommand{\arraystretch}{0.85}\begin{tabular}[c]{@{}c@{}}E\end{tabular}} & \cellcolor{grpa5699a3f} {\renewcommand{\arraystretch}{0.85}\begin{tabular}[c]{@{}c@{}}\ensuremath{\checkmark}\end{tabular}} & \cellcolor{grpa5699a3f} {\renewcommand{\arraystretch}{0.85}\begin{tabular}[c]{@{}c@{}}\ensuremath{\checkmark}\end{tabular}} \\
\cellcolor{grpa5699a3f}  & \cellcolor{grpa5699a3f} {\renewcommand{\arraystretch}{0.85}\begin{tabular}[c]{@{}l@{}}Use Column-Pair Stats\\[-0.4ex]for Correlated Filters\end{tabular}} & \cellcolor{grpa5699a3f} {\renewcommand{\arraystretch}{0.85}\begin{tabular}[c]{@{}c@{}}E\end{tabular}} & \cellcolor{grpa5699a3f} {\renewcommand{\arraystretch}{0.85}\begin{tabular}[c]{@{}c@{}}--\end{tabular}} & \cellcolor{grpa5699a3f} {\renewcommand{\arraystretch}{0.85}\begin{tabular}[c]{@{}c@{}}\ensuremath{\checkmark}\end{tabular}} \\
\cellcolor{grpa5699a3f}  & \cellcolor{grpa5699a3f} {\renewcommand{\arraystretch}{0.85}\begin{tabular}[c]{@{}l@{}}Equality from Frequent\\[-0.4ex]Values or Rare Tail\end{tabular}} & \cellcolor{grpa5699a3f} {\renewcommand{\arraystretch}{0.85}\begin{tabular}[c]{@{}c@{}}C\end{tabular}} & \cellcolor{grpa5699a3f} {\renewcommand{\arraystretch}{0.85}\begin{tabular}[c]{@{}c@{}}39\%\end{tabular}} & \cellcolor{grpa5699a3f} {\renewcommand{\arraystretch}{0.85}\begin{tabular}[c]{@{}c@{}}31\%~preds\end{tabular}} \\
\cellcolor{grpa5699a3f}  & \cellcolor{grpa5699a3f} {\renewcommand{\arraystretch}{0.85}\begin{tabular}[c]{@{}l@{}}LIKE Patterns from\\[-0.4ex]Stored Substrings\end{tabular}} & \cellcolor{grpa5699a3f} {\renewcommand{\arraystretch}{0.85}\begin{tabular}[c]{@{}c@{}}E\end{tabular}} & \cellcolor{grpa5699a3f} {\renewcommand{\arraystretch}{0.85}\begin{tabular}[c]{@{}c@{}}23\%\end{tabular}} & \cellcolor{grpa5699a3f} {\renewcommand{\arraystretch}{0.85}\begin{tabular}[c]{@{}c@{}}24\%~preds\end{tabular}} \\
\cellcolor{grpa5699a3f}  & \cellcolor{grpa5699a3f} {\renewcommand{\arraystretch}{0.85}\begin{tabular}[c]{@{}l@{}}Range Filters from\\[-0.4ex]Numeric Buckets\end{tabular}} & \cellcolor{grpa5699a3f} {\renewcommand{\arraystretch}{0.85}\begin{tabular}[c]{@{}c@{}}C\end{tabular}} & \cellcolor{grpa5699a3f} {\renewcommand{\arraystretch}{0.85}\begin{tabular}[c]{@{}c@{}}14\%\end{tabular}} & \cellcolor{grpa5699a3f} {\renewcommand{\arraystretch}{0.85}\begin{tabular}[c]{@{}c@{}}12\%~preds\end{tabular}} \\
\cellcolor{grpa5699a3f}\multirow{-8}{*}{\rotatebox[origin=c]{90}{{\renewcommand{\arraystretch}{0.85}\begin{tabular}[c]{@{}c@{}}Filter\\[-0.4ex]Selectivity\end{tabular}}}} & \cellcolor{grpa5699a3f} {\renewcommand{\arraystretch}{0.85}\begin{tabular}[c]{@{}l@{}}Estimate Predicates\\[-0.4ex]Comparing Two Columns\end{tabular}} & \cellcolor{grpa5699a3f} {\renewcommand{\arraystretch}{0.85}\begin{tabular}[c]{@{}c@{}}E\end{tabular}} & \cellcolor{grpa5699a3f} {\renewcommand{\arraystretch}{0.85}\begin{tabular}[c]{@{}c@{}}--\end{tabular}} & \cellcolor{grpa5699a3f} {\renewcommand{\arraystretch}{0.85}\begin{tabular}[c]{@{}c@{}}4\%~preds\end{tabular}} \\
\hline
\cellcolor{grpa0fcf967}  & \cellcolor{grpa0fcf967} {\renewcommand{\arraystretch}{0.85}\begin{tabular}[c]{@{}l@{}}Join from Overlap\\[-0.4ex]of Frequent Keys\end{tabular}} & \cellcolor{grpa0fcf967} {\renewcommand{\arraystretch}{0.85}\begin{tabular}[c]{@{}c@{}}E\end{tabular}} & \cellcolor{grpa0fcf967} {\renewcommand{\arraystretch}{0.85}\begin{tabular}[c]{@{}c@{}}--\end{tabular}} & \cellcolor{grpa0fcf967} {\renewcommand{\arraystretch}{0.85}\begin{tabular}[c]{@{}c@{}}\ensuremath{\checkmark}\end{tabular}} \\
\cellcolor{grpa0fcf967}  & \cellcolor{grpa0fcf967} {\renewcommand{\arraystretch}{0.85}\begin{tabular}[c]{@{}l@{}}Join from Shared Keys\\[-0.4ex]and Rows per Key\end{tabular}} & \cellcolor{grpa0fcf967} {\renewcommand{\arraystretch}{0.85}\begin{tabular}[c]{@{}c@{}}E\end{tabular}} & \cellcolor{grpa0fcf967} {\renewcommand{\arraystretch}{0.85}\begin{tabular}[c]{@{}c@{}}\ensuremath{\checkmark}\end{tabular}} & \cellcolor{grpa0fcf967} {\renewcommand{\arraystretch}{0.85}\begin{tabular}[c]{@{}c@{}}\ensuremath{\checkmark}\end{tabular}} \\
\cellcolor{grpa0fcf967}  & \cellcolor{grpa0fcf967} {\renewcommand{\arraystretch}{0.85}\begin{tabular}[c]{@{}l@{}}Treat Join Chains\\[-0.4ex]as One Shared-Key Group\end{tabular}} & \cellcolor{grpa0fcf967} {\renewcommand{\arraystretch}{0.85}\begin{tabular}[c]{@{}c@{}}C\end{tabular}} & \cellcolor{grpa0fcf967} {\renewcommand{\arraystretch}{0.85}\begin{tabular}[c]{@{}c@{}}--\end{tabular}} & \cellcolor{grpa0fcf967} {\renewcommand{\arraystretch}{0.85}\begin{tabular}[c]{@{}c@{}}\ensuremath{\checkmark}\end{tabular}} \\
\cellcolor{grpa0fcf967}  & \cellcolor{grpa0fcf967} {\renewcommand{\arraystretch}{0.85}\begin{tabular}[c]{@{}l@{}}Shared-Key Overlap across\\[-0.4ex]Three or More Tables\end{tabular}} & \cellcolor{grpa0fcf967} {\renewcommand{\arraystretch}{0.85}\begin{tabular}[c]{@{}c@{}}E\end{tabular}} & \cellcolor{grpa0fcf967} {\renewcommand{\arraystretch}{0.85}\begin{tabular}[c]{@{}c@{}}--\end{tabular}} & \cellcolor{grpa0fcf967} {\renewcommand{\arraystretch}{0.85}\begin{tabular}[c]{@{}c@{}}\ensuremath{\checkmark}\end{tabular}} \\
\cellcolor{grpa0fcf967}  & \cellcolor{grpa0fcf967} {\renewcommand{\arraystretch}{0.85}\begin{tabular}[c]{@{}l@{}}Shrink Distinct-Key Counts\\[-0.4ex]after Filters\end{tabular}} & \cellcolor{grpa0fcf967} {\renewcommand{\arraystretch}{0.85}\begin{tabular}[c]{@{}c@{}}E\end{tabular}} & \cellcolor{grpa0fcf967} {\renewcommand{\arraystretch}{0.85}\begin{tabular}[c]{@{}c@{}}\ensuremath{\checkmark}\end{tabular}} & \cellcolor{grpa0fcf967} {\renewcommand{\arraystretch}{0.85}\begin{tabular}[c]{@{}c@{}}\ensuremath{\checkmark}\end{tabular}} \\
\cellcolor{grpa0fcf967}  & \cellcolor{grpa0fcf967} {\renewcommand{\arraystretch}{0.85}\begin{tabular}[c]{@{}l@{}}Push Filtered Dimension IDs\\[-0.4ex]into Fact Tables\end{tabular}} & \cellcolor{grpa0fcf967} {\renewcommand{\arraystretch}{0.85}\begin{tabular}[c]{@{}c@{}}E\end{tabular}} & \cellcolor{grpa0fcf967} {\renewcommand{\arraystretch}{0.85}\begin{tabular}[c]{@{}c@{}}--\end{tabular}} & \cellcolor{grpa0fcf967} {\renewcommand{\arraystretch}{0.85}\begin{tabular}[c]{@{}c@{}}\ensuremath{\checkmark}\end{tabular}} \\
\cellcolor{grpa0fcf967}  & \cellcolor{grpa0fcf967} {\renewcommand{\arraystretch}{0.85}\begin{tabular}[c]{@{}l@{}}Foreign-Key Rows Matched\\[-0.4ex]to Filtered Parents\end{tabular}} & \cellcolor{grpa0fcf967} {\renewcommand{\arraystretch}{0.85}\begin{tabular}[c]{@{}c@{}}C\end{tabular}} & \cellcolor{grpa0fcf967} {\renewcommand{\arraystretch}{0.85}\begin{tabular}[c]{@{}c@{}}65\%\end{tabular}} & \cellcolor{grpa0fcf967} {\renewcommand{\arraystretch}{0.85}\begin{tabular}[c]{@{}c@{}}54\%~joins\end{tabular}} \\
\cellcolor{grpa0fcf967}\multirow{-8}{*}{\rotatebox[origin=c]{90}{{\renewcommand{\arraystretch}{0.85}\begin{tabular}[c]{@{}c@{}}Join\\[-0.4ex]Cardinality\end{tabular}}}} & \cellcolor{grpa0fcf967} {\renewcommand{\arraystretch}{0.85}\begin{tabular}[c]{@{}l@{}}Movie-ID Star-Schema\\[-0.4ex]Join Model\end{tabular}} & \cellcolor{grpa0fcf967} {\renewcommand{\arraystretch}{0.85}\begin{tabular}[c]{@{}c@{}}E\end{tabular}} & \cellcolor{grpa0fcf967} {\renewcommand{\arraystretch}{0.85}\begin{tabular}[c]{@{}c@{}}60\%\end{tabular}} & \cellcolor{grpa0fcf967} {\renewcommand{\arraystretch}{0.85}\begin{tabular}[c]{@{}c@{}}59\%~joins\end{tabular}} \\
\hline
\end{tabular}
  }
  \caption{Overview of strategies employed in the separate cardinality estimators synthesized for JOB and JOB-Complex. 
  The strategies are classified into classical (C) and exotic (E). 
  Classical ones can also be found in general-purpose cardinality estimators, but here they are used in a workload-specific manner. 
  Exotic strategies are not commonly found in general-purpose estimators.}
  \label{tab:semantic_eval}
  \vspace{-5ex}
\end{table}

In this subsection we inspect the estimators synthesized by \system{} to understand where the improvements come from\footnote{The synthesized estimators can be found \textit{\href{https://github.com/DataManagementLab/BespokeCard/tree/main/synthesized\_card\_estimators}{here}}.}.

The strategy breakdown in \Cref{tab:semantic_eval} summarizes the same picture quantitatively: the generated estimators are not just PostgreSQL-style catalogs with different constants, but executable strategies specialized to the IMDb schema and JOB/JOB-Complex workload.
This is also the empirical evidence for the architectural argument of \Cref{sec:background}: the two estimators mix histograms, samples, trigram and prefix summaries, per-\texttt{info\_type\_id} conditional statistics, PK/FK fanout logic, cached string patterns, and pair statistics within a single artifact, a combination that no fixed traditional or learned model family would commit to in advance.

\subsubsection{Synthesized statistics.}
\noindent\textbf{Familiar ingredients as the base.}
Both estimators start from familiar ingredients: table cardinalities, null fractions, distinct-value counts, top-$k$ value summaries, and histograms for ordered integer columns such as production years, episode numbers, and ordering attributes.
The difference is where this budget is spent.
\system{} assigns large samples and richer summaries to high-impact fact tables such as \texttt{cast\_info} and \texttt{movie\_info}, while tiny dimensions such as \texttt{kind\_type} and \texttt{role\_type} are summarized nearly exactly.
For string-heavy attributes in titles, names, companies, and keywords, it combines exact top-$k$ values, prefix frequencies, and trigram summaries.
For recurring note predicates, it builds token-level document frequencies for workload fragments such as \texttt{voice}, \texttt{producer}, \texttt{uncredited}, \texttt{USA}, and \texttt{worldwide}.

\noindent\textbf{Schema-specific summaries.}
The most schema-specific statistics concern the overloaded IMDb information tables.
The \texttt{info} columns in IMDb information tables do not represent one homogeneous domain: their meaning depends on \texttt{info\_type\_id}, which distinguishes genres, countries, certificates, runtimes, release notes, and other categories.
The synthesized estimators therefore build separate sub-statistics per \texttt{info\_type\_id}: categorical summaries for low-cardinality groups such as genres and countries, text summaries for larger groups, and lexicographic histograms for comparisons in \texttt{movie\_info\_idx}.
This is a concrete example of \system{} adapting to schema semantics rather than treating a physical column as one undifferentiated distribution.

\noindent\textbf{Correlations and string patterns for JOB-Complex.}
JOB-Complex receives a broader design because its predicates and joins are less regular.
In addition to the basic summaries, it synthesizes top-$k$ statistics over phonetic-code columns, IMDb-index columns, notes, and non-key attributes.
It also adds explicit pair statistics for correlated columns, for example title kind/year, movie-information type/value, and cast-person note pairs.
These pair summaries let the estimator avoid multiplying independent marginals when the workload exposes correlated predicates.
For string predicates, JOB-Complex goes even further: it caches concrete workload patterns such as \texttt{\%Warner\%}, \texttt{Lionsgate\%}, \texttt{\%Downey\%}, \texttt{Saw\%}, \texttt{\%(USA)\%}, \texttt{USA:\% 199\%}, and \texttt{\%follow\%}.
These literals are too workload-specific for a general DBMS catalog, but they are exactly the kind of information a bespoke estimator can exploit.

\subsubsection{Synthesized estimation logic}
The generated code turns these statistics into estimation routines for predicates and joins.
For base-table filters, equality and \texttt{IN} predicates use exact counts or top-$k$ frequencies with explicit tail handling; ranges use histograms or value maps; and \texttt{LIKE} predicates use exact evaluation for tiny tables, cached counts for known workload patterns, prefix statistics for anchored patterns, and trigram/example-based fallbacks for unanchored substrings.
Boolean expressions are handled recursively, with conjunctions multiplied and disjunctions combined using inclusion-exclusion-style formulas.

\noindent\textbf{Schema-aware join estimation.}
For joins, the estimators use the IMDb schema rather than only applying generic join-cardinality estimations formulas like:
$|R \bowtie S| \approx \frac{|R| \cdot |S|}{\max(\mathrm{ndv}(R.a), \mathrm{ndv}(S.b))}$.
Both synthesized estimators contain an explicit PK/FK map, preserve uniqueness for primary keys, track effective NDVs after filtering, and propagate filtered key fractions from dimensions into fact tables.
Thus a filter on \texttt{kind\_type} before joining into \texttt{title} is not treated like an arbitrary overlap between two large relations.

The JOB estimator is especially shaped around the movie domain.
It recognizes \texttt{title.id} and the many \texttt{movie\_id} columns as a shared key space, estimates coverage and average rows per movie for each movie-indexed table, and propagates this information through joins.
This encodes the fact that joining \texttt{cast\_info}, \texttt{movie\_companies}, \texttt{movie\_keyword}, and \texttt{movie\_info} is a collection of fanouts around the same movie universe, not a generic many-way join over unrelated keys.

\noindent\textbf{Broader machinery for JOB-Complex.}
The JOB-Complex estimator adds machinery for the harder cases.
It performs exact pushdown from tiny dimensions such as \texttt{info\_type}, \texttt{kind\_type}, \texttt{company\_type}, \texttt{link\_type}, \texttt{role\_type}, and \texttt{comp\_cast\_type}: a semantic filter on a dimension label is evaluated directly and converted into an \texttt{IN} restriction on the corresponding foreign key.
It also uses pair statistics to override independence, for example estimating a conjunction over \texttt{info\_type\_id} and \texttt{info} in \texttt{movie\_info} from their joint summary.
For joins beyond PK/FK structure, it builds join equivalence classes, distinguishes movie-star joins from generic overlap joins, uses top-$k$ overlap for string/code-based joins when available, and conservative tail estimates otherwise.

\noindent\textbf{Adapting to different workloads.}
JOB leads \system{} toward large samples, conditional information-table statistics, note-token summaries, PK/FK metadata, and movie-domain fanout logic.
JOB-Complex leads it toward more distributed summaries, pair statistics, cached string patterns, exact pushdown from tiny dimensions, column-equality handling, and more general join logic.
The contrast in the two synthesized estimators outlines the capabilties of \system{} to adapt to different workloads, exploiting vastly different strategies to achieve speedups.
Rather than merely tuning a fixed estimator, \system{} produces an estimator structure that suit the database and workload.

\vspace{-.5ex}
\subsection{Impact of Staged Feedback}
\label{sec:eval-stages}

\begin{figure}
  \centering
  \includegraphics[width=\linewidth]{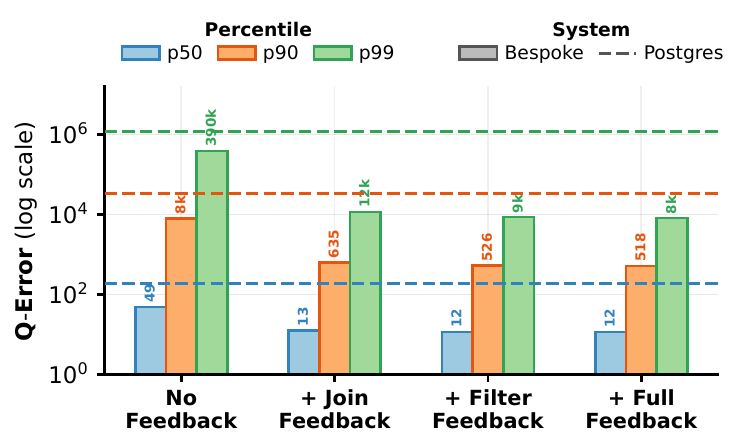}
  \vspace{-5ex}
  \caption{Q-error distribution over all subplans of the JOB workload after each stage of the synthesis loop. The initial estimator, produced without any feedback, already outperforms PostgreSQL. Each subsequent feedback stage widens the gap, eventually by orders of magnitude.}
  \label{fig:q_error_by_stage}
  \vspace{-2ex}
\end{figure}

A central design choice of \system{} is the staged feedback loop of \Cref{subsec:feedback}.
To quantify its contribution, \Cref{fig:q_error_by_stage} reports the q-error percentiles over all JOB subplans after each stage of synthesis.

The initial estimator design and implementation, produced without any empirical feedback, is the starting point of the synthesis loop.
Even this feedback-free estimator already outperforms PostgreSQL: it reaches a median q-error of 48.5 on all JOB subplans, compared to PostgreSQL's 190.
Each subsequent feedback stage further reduces the error: the median falls to 11.5 after the final stage, and the upper percentiles improve dramatically, with the 90th and 95th percentiles ending at 517 and 1k versus PostgreSQL's 33k and 1.1m.
This demonstrates two complementary findings.
First, the staged feedback loop is effective: it repairs systematic failure modes that the initial design misses, improving accuracy by orders of magnitude at the tail.
Second, the approach is robust even without extensive optimization, since the feedback-free estimator alone already improves substantially over PostgreSQL, highlighting the strength of the planner/coder decomposition itself.

\vspace{-.5ex}
\subsection{Storage Footprint}
\label{sec:eval-memory}

\begin{table}[]
  \centering
  \begin{tabular}{lrr}
    \toprule
    \textbf{Workload} & \textbf{Initial Impl.} & \textbf{Final Optimization} \\
    \midrule
    JOB & 32.2\,MB & 31.4\,MB \\
    JOB-Complex & 17.9\,MB & 26.0\,MB \\
    \bottomrule
  \end{tabular}
  \caption{Storage footprint of the synthesized statistics for the initial implementation and the final optimized estimator (measured as python memory consumption). The footprint is small relative to the full dataset (IMDB ~3.6\,GB).}
  \label{tab:storage_footprint}
  \vspace{-4ex}
\end{table}

Because the estimator must serve as a lightweight optimizer component, the size of the statistics it builds matters.
\Cref{tab:storage_footprint} reports the storage footprint of the created statistics (measured as Python memory consumption) for the initial implementation and the final optimized estimator.
Overall, the footprint is small, with the initial estimator occupying only 32.2\,MB on JOB and 17.9\,MB on JOB-Complex, roughly 0.9\% and 0.5\% of the $\sim$3.6\,GB IMDB dataset.
The feedback loop does not inflate this budget: on JOB the footprint even shrinks slightly to 31.4\,MB, while on JOB-Complex it grows modestly to 26.0\,MB (about 0.7\% of the dataset) as the coder adds statistics to address residual errors.
This is consistent with the \texttt{estimator\_size} feedback (\Cref{tab:feedback-fields}), which discourages buying accuracy with unbounded statistics, and the absolute numbers could be reduced further with a more memory-efficient programming language than our Python prototype.

\vspace{-.5ex}
\subsection{Synthesis Cost \& Code Size}
\label{sec:eval-cost}

\begin{figure}
  \centering
  \includegraphics[width=\linewidth]{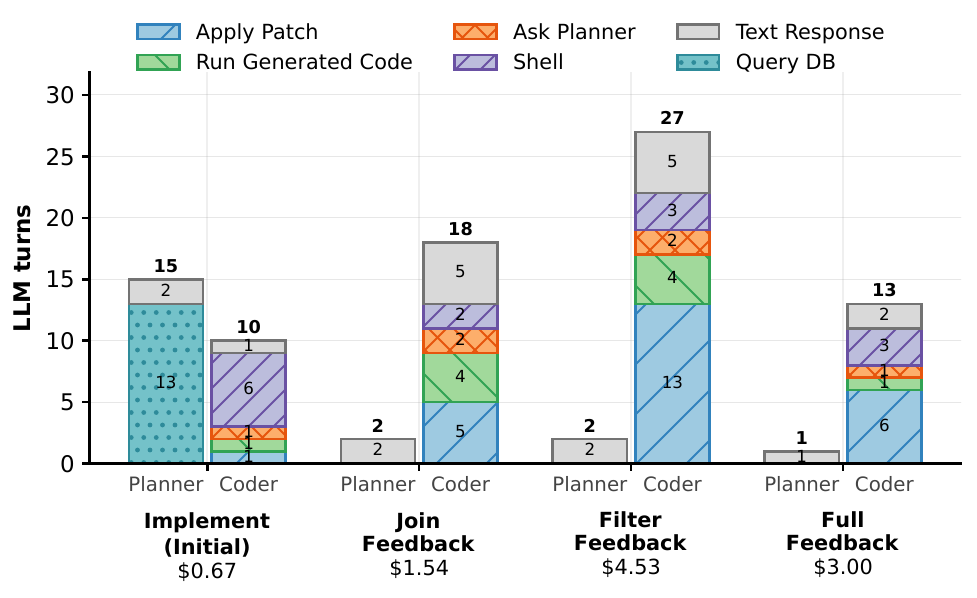}
  \vspace{-5ex}
  \caption{Number of agent turns, grouped by action type, for the planner and coder agents per synthesis stage on the JOB workload. The complete run takes 88 turns and costs \$9.74.}
  \label{fig:calls_by_stage}
  \vspace{-4ex}
\end{figure}

Finally, we report the cost of running \system{} itself.
\Cref{fig:calls_by_stage} breaks down the agent turns made by the planner and coder across the synthesis stages on the JOB workload, counting both tool-using turns (database queries, patch application, shell checks, and generated-code runs) and text-only responses, since both consume LLM context and latency.
The complete JOB run takes 88 agent turns using GPT-5.4 and costs \$9.74: \$0.67 for the initial implementation, \$1.54 for join-feedback repair, \$4.53 for filter-feedback repair, and \$3.00 for the final full-feedback stage.
The planner accounts for only 20 turns, most of them before the first implementation, where it spends 13 database-query turns and two text responses to inspect the workload and design the initial statistics plan; later stages require only two, two, and one planner turns.

The coder performs the remaining 68 turns of implementation and repair work, applying 25 patches, executing the generated estimator 10 times, issuing 14 shell checks, and asking the planner six clarification questions.
The most expensive stage is filter-feedback repair (29 turns, \$4.53), because string and base-table predicate errors require more localized code changes than the initial join repair.
Even so, producing the fully optimized estimator completes in under one hour for less than \$10.

The produced estimator code for JOB contains 1.3k lines of code, and 1.2k lines for JOB-Complex. 
Despite its modest size, this code encapsulates both the statistics-generation logic and the workload-specific estimation strategies synthesized by \system{}, including specialized handling of joins, filters, correlations, and fallback cases.
Rather than generating a large software system, \system{} converges on concise workload-specific estimators whose complexity is driven by the structure of the workload rather than by a fixed estimator architecture.
The modest size of the estimators also allows inspection, auditing and modification of the estimator by human engineers if requested.
\vspace{-1ex}
\section{Conclusion and Future Work}
\vspace{-.5ex}
\label{sec:conclusion}

\system{} reframes cardinality estimation as the synthesis of an \emph{executable} estimator specialized to a fixed database and tuned for a workload, rather than the configuration of a generic statistics catalog or the training of a predetermined learned model.
Given a database and a representative workload, \system{} pairs a planning agent that designs workload-specific statistics and estimation strategies with a coding agent that implements them as runnable code, and closes the loop with a deterministic evaluator that turns subplan-level q-errors and regressions against PostgreSQL into structured improvement signals.
A staged curriculum over join-only, filter-only, and full-subplan objectives localizes estimation failures and lets the loop repair them incrementally.
The resulting artifact is an inspectable, repairable, and cheaply deployable estimator whose statistics and estimation logic are tailored to the declared database-and-workload contract.

Our evaluation on JOB and JOB-Complex over the IMDB dataset shows that this approach yields accuracy gains that translate directly into faster execution.
Injecting \system{}'s estimates into PostgreSQL's optimizer reduces total runtime by 33\% on JOB and by 68\% on JOB-Complex, closing the gap to the unattainable true-cardinality ceiling to within 15\% and 9\% of PostgreSQL's original runtime.
These gains stem from estimates that are more accurate at every percentile: \system{} lowers the median q-error and, more importantly, compresses the heavy error tail that drives the optimizer's worst plan choices.
On JOB-Complex the 95th-percentile q-error falls from 590k to 3k.
Two findings stand out.
First, even the feedback-free initial estimator already outperforms PostgreSQL, demonstrating the strength of the planner/coder decomposition on its own.
Second, the staged feedback loop widens this gap by orders of magnitude at the tail.
\system{} achieves all of this with a small footprint (about 31\,MB, under 1\% of the database size) and a modest synthesis budget of fewer than 100 GPT-5.4 LLM calls, completing in under one hour for less than \$10 per workload.

The current prototype specializes to a declared workload and database snapshot, so shifting data distributions, new query templates, or broader SQL coverage may require refreshing statistics or resynthesizing the estimator.
These constraints suggest several directions for future research.
A first direction is incremental maintainability: rather than resynthesizing from scratch when the database evolves, the loop could detect distribution drift and repair only the affected statistics and estimation logic, treating updateability as an additional feedback signal and raising the question of how to characterize drift cheaply and decide when local repair suffices versus full resynthesis.
A second direction concerns deployment efficiency: our estimators are Python prototypes, and porting the generated statistics and estimation logic to a compiled language (potentially by having the synthesis loop itself target a compiled backend) would shrink both memory footprint and inference latency, making the synthesized estimators viable as in-process optimizer components without sacrificing inspectability.
A third direction is generality across optimizer components, since the feedback-driven synthesis pattern is not specific to cardinality estimation: cost models, hint-selection policies, and other components are likewise governed by a database-and-workload contract and admit deterministic, measurable feedback, so they could be synthesized within the same loop, and jointly synthesizing several interacting components, while understanding how their feedback signals compose, points toward optimizers assembled entirely from workload-specialized, synthesized code.
More broadly, we believe that treating optimizer components as code to be synthesized under structured feedback, rather than as fixed models to be configured, is a promising direction for building database systems that adapt to the data and queries they actually serve.

\bibliographystyle{ACM-Reference-Format}
\bibliography{bib}

\end{document}